\documentclass[nofootinbib,preprint,noshowpacs,noshowkeys,floatfix,aps]{revtex4-1}
\usepackage{bm}
\usepackage{graphicx} 
\usepackage{amsmath}
\usepackage{mathtools}
\usepackage{booktabs}
\usepackage{tabularx}
\usepackage[thinc]{esdiff}
\usepackage{threeparttable}
\usepackage{flafter}
\usepackage{xcolor}
\usepackage{balance}
\usepackage{breakurl}
\usepackage{xurl}
\usepackage{url}

\newcommand{\tj}[6]{ \begin{pmatrix}
  #1 & #2 & #3 \\
  #4 & #5 & #6 
 \end{pmatrix}}
 
 \newcommand{\Gj}[6]{ \begin{Bmatrix}
  #1 & #2 & #3 \\
  #4 & #5 & #6 
 \end{Bmatrix}}

\begin{document}
\title{Dynamics of translational and rotational thermalization \\ of AlF molecules via collisions with cryogenic helium}

\author{M. Karra}
\affiliation{Fritz-Haber-Institut der Max-Planck-Gesellschaft, Faradayweg 4-6, 14195 Berlin, Germany }

\author{M. T. Cretu}
\affiliation{Fritz-Haber-Institut der Max-Planck-Gesellschaft, Faradayweg 4-6, 14195 Berlin, Germany }

\author{B. Friedrich}
\affiliation{Fritz-Haber-Institut der Max-Planck-Gesellschaft, Faradayweg 4-6, 14195 Berlin, Germany }

\author{S. Truppe}
\affiliation{Fritz-Haber-Institut der Max-Planck-Gesellschaft, Faradayweg 4-6, 14195 Berlin, Germany }

\author{G. Meijer}
\affiliation{Fritz-Haber-Institut der Max-Planck-Gesellschaft, Faradayweg 4-6, 14195 Berlin, Germany }

\author{J. P\'erez-R\'ios}
\affiliation{Fritz-Haber-Institut der Max-Planck-Gesellschaft, Faradayweg 4-6, 14195 Berlin, Germany }

\date{\today}
\begin{abstract}
We investigated helium-mediated translational and rotational thermalization of the aluminum monofluoride (AlF) molecule at cryogenic temperatures via a new \emph{ab initio} potential energy surface (PES) and quantum multichannel scattering theory. Our examination of the elastic and rotationally inelastic channels revealed that helium is an efficient quencher of AlF at temperatures relevant to buffer gas cooling experiments ($\sim1$~mK to $10$~K). We also showed that this conclusion is robust against possible inaccuracies of the PES.


\end{abstract}

\maketitle

\section{Introduction}

The ability to cool gaseous ensembles of molecules to cold ($1$~mK to $10$ K) and ultracold ($\lesssim$1~mK) temperatures has been impacting research areas as diverse as testing the standard model of particle physics~\cite{Safronova2018}, quantum-logic spectroscopy~\cite{Mur2012,Shi2013,Wolf2016}, action-spectroscopy~\cite{Hudson2014,Sandra2017,Sandra2020}, chemical reaction dynamics~\cite{Krems2004,Krems2008,Carr2009,Schnell2009,Quemener2012,Bala2016,JPRBook}, as well as quantum computing ~\cite{DeMille,Ketan,Yu_2019} and quantum simulation ~\cite{Madison2013,Blackmore2018}. 
In particular, ultracold \emph{polar} molecules afford tunability of interactions, coveted for applications in quantum information protocols~\cite{Zhu2013,Sawant2020} and the simulation of many-body Hamiltonians~\cite{Micheli2006,Herrera2011}. 

Ultracold molecules can be produced via direct or indirect cooling techniques. The \emph{indirect} techniques rely on photo-association~\cite{Fioretti1998,Bohn1999,Jones2006,JPR2015,Blasing2016} or magneto-association~\cite{Kohler2006,Chin2010} of a pair of ultracold atoms and have proved especially successful for generating bi-alkali ultracold molecules~\cite{Ni2008,Danzl2008,Ye2018,DeMarco2019,Yang2019,Voges2020}. The yield of the indirect techniques is limited by the number of available ultracold atoms that serve as precursors of the ultracold molecules. However, bi-alkali molecules are prone to undergoing sticky collisions: these correspond to the formation of an intermediate complex whose lifetime can be long enough to amount to a loss of the molecules formed~\cite{Mayle2012,Sticky2019,Gregory2019}. 

On the other hand, \emph{direct} cooling techniques are based on the dissipation of translational -- and internal -- energy of pre-existing molecules via thermalization with a cryogenic buffer gas~\cite{Weinstein_1998}, Stark~\cite{Bethlem_1999} or Zeeman~\cite{Vanhaecke_2007,Narevicius_2008} slowing of molecules cooled by a supersonic expansion, Sisyphus cooling \cite{Zeppenfeld_2012} or, most recently, laser cooling \cite{Tarbutt_2018,Hudson_2020}. Long considered impractical if not impossible~\cite{EPJD2004}, laser cooling of molecules has led, for the first time, to the creation of ultracold molecular ensembles \cite{Mitra_2020,Augenbraun_2020,AugenbraunPRX_2020}.  As laser cooling of molecules requires near-unity Franck-Condon factors between the molecular ground and excited electronic states~\cite{Rosa2004,Shuman2010}, only a handful of molecules have been laser-cooled so far~\cite{Ivanov2019,osti_1642256,Augenbraun2020,Klos2020}.

  
Recently, we identified the aluminum monofluoride (AlF) molecule as a promising candidate for laser cooling. Its virtues include a large photon scattering rate and favorable Franck-Condon factors~\cite{Truppe2019,hofsass2021optical,Doppelbauer2020}. Moreover, we demonstrated that AlF molecules can be efficiently produced in a pulsed buffer-gas source, with a yield of $\gtrsim 10^{12}$ molecules per steradian per pulse~\cite{hofsass2021optical}. A high yield of cryogenically thermalized molecules is a key prerequisite for reaching quantum degeneracy in a cooling step that follows upon laser cooling, namely forced evaporation.

In order to gain insight into the thermalization of AlF by cryogenic helium, we undertook a study of the quantum dynamics of the elastic and rotationally inelastic scattering of AlF by He at collision energies ranging from $1$ mK to $10$ K. We made use of a new and accurate {\it ab initio} potential energy surface (PES) and  calculated the relevant scattering observables via the coupled-channel method. The hyperfine structure of AlF has been omitted as it is expected to play only a minor role in the collision dynamics. We also discuss what our findings about the AlF-He system imply about the thermalization in a buffer gas of other $^1\Sigma^+$ diatomic fluorides, such as MgF and BaF. While it is known that AlF forms in highly vibrationally excited levels\cite{Rosenwaks1976}, and that the vibrational quenching also needs to be treated in detail, it is beyond the scope of the present work and left for a future article.

We further note that, following the discovery of its presence in the envelope of proto-planetary nebulae~\cite{Highberger2001}, the AlF molecule and its spectroscopic properties have been also of astronomical interest. More recently, the $\gamma$-ray emission from $^{26}\textrm{AlF}$ has served as a stepping stone to elucidating the Galactic sources of $^{26}\textrm{Al}$ \cite{AlF_radioactive}, which is of consequence for understanding the frequency of core collapse of supernovae \cite{26Al}. 

\section{Potential energy surface \label{Sec.PES}}

\begin{figure}
\centering
\includegraphics[width=1\linewidth]{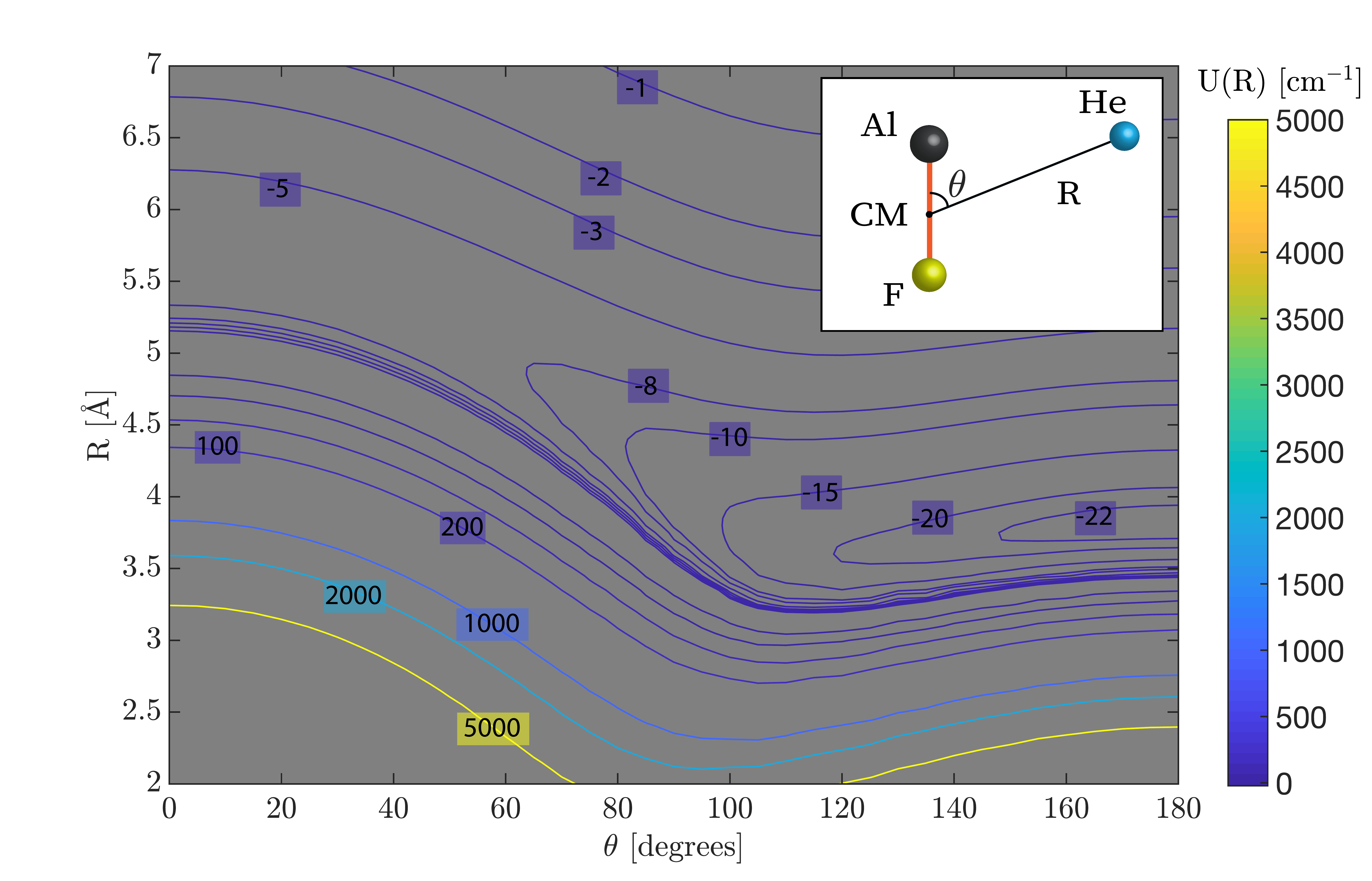}
\caption{Potential energy surface (PES) of the AlF(X$^1\Sigma^+$)--He($^1S$) system. The contours show fixed electronic energies in cm$^{-1}$ of the system as a function of $R$ and $\theta$, whose meaning is defined in the inset.  See also text.}
\label{fig1}
\end{figure}

The potential energy surface of the $^{27}$Al$^{19}$F(X$^1\Sigma^+$)+$^4$He(1$S$) system was calculated by treating AlF(X$^1\Sigma^+$) like a vibrationless rigid rotor. Thus, the PES only depends on two parameters: the distance $R=|\bf{R}|$ between the center of mass of the molecule and the atom and the angle $\theta$ between $\bf{R}$ and the internuclear axis $\bf{r}$ of the AlF molecule, see the inset in Fig.~\ref{fig1}. The PES was calculated at the coupled-cluster level of theory, including singles and doubles excitations; triples excitations were treated perturbatively [RCCSD(T)] as implemented in Molpro~\cite{MOLPRO} using the aug-cc-pv5Z basis set of Dunning~\cite{BasisSet}. The Al-F equilibrium distance was fixed to that accurately determined in a recent experiment \cite{Truppe2019,database}, $R_e=1.654369$ $\textrm{\AA}$. The superposition error was corrected using the counterpoise method. A contour plot of the PES is shown in Fig.~\ref{fig1}. Note that the AlF-He attraction is quite weak, as it arises from the van der Waals interaction between a closed-shell atom and a closed-shell molecule. The PES exhibits a global minimum of $\sim 22$~cm$^{-1}$ close to the F atom, lending it a high anisotropy characteristic of fluorine-containing heteronuclear  molecules interacting with He~\cite{YbFHe}.

The PES was calculated for 1500 geometries including 20 angles, 18 of which corresponded to the Gauss-Legendre quadrature points and to $\theta=0$ and $\theta= \pi$, and 75 radial points between 1.6 and 30~$\textrm{\AA}$ with a larger density of points for $R$ between 3 and 5 $\textrm{\AA}$. The global minimum of the PES was found at $\theta= \pi$ and $R=3.8~\textrm{\AA}$, in close agreement with previously reported calculations~\cite{AlFHe_previous}. The technical details of calculating the PES will be published elsewhere. 

\begin{figure}
\centering
\includegraphics[width=1\linewidth]{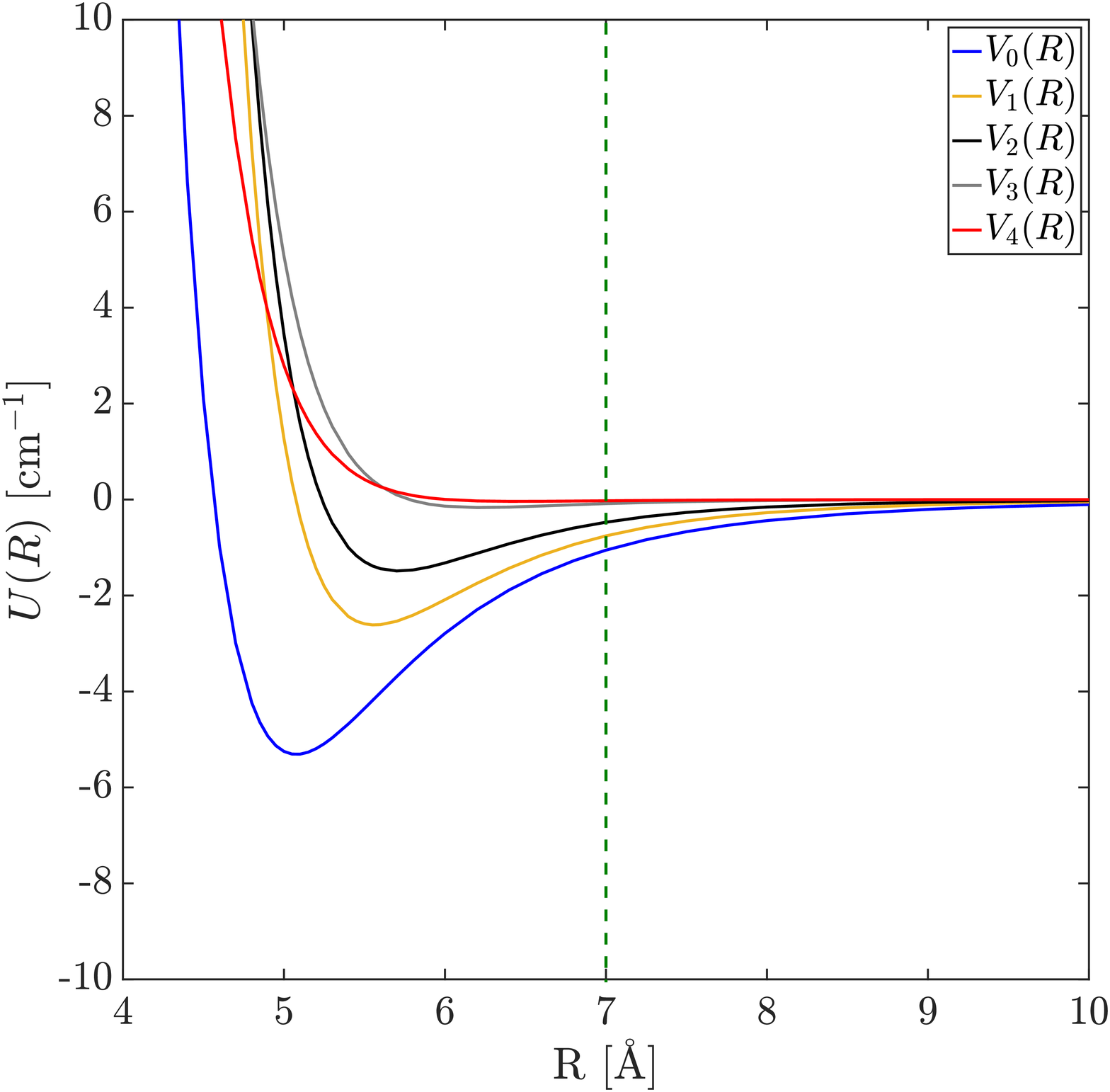}
\caption{The first five radial terms $V_{\lambda}(R)$ of the potential energy surface of the AlF(X$^1\Sigma^+$)-He(1$S$) system, see text. 
}
\label{fig2}
\end{figure}

For scattering calculations, it is advantageous to expand the PES in a Legendre series
\begin{equation}
\label{eq_PES}
V(R,\theta)=\sum_{\lambda=0}^{\lambda_\text{max}}V_{\lambda}(R)P_\lambda(\cos{\theta})    
\end{equation}
\noindent
with $P_\lambda(\cos \theta)$ Legendre polynomials of degree $\lambda$ and expansion coefficients 
\begin{equation}
\label{eq_vlambda}
    V_{\lambda}(R)=\left(\frac{2\lambda+1}{2} \right)\int_{0}^{\pi} V(R,\theta)P_\lambda(\cos{\theta}) \sin{\theta}d\theta
\end{equation}
that depend solely on the radial distance $R$. Note that $V_0(R)$ represents the only isotropic (spherical) contribution to the PES whereas all other contributions are anisotropic. The integral in Eq.~(\ref{eq_vlambda}) was performed using the Gauss-Legendre quadrature scheme employing 16 quadrature points. The dependence of the first five $V_{\lambda}(R)$ terms on $R$ is shown in Fig.~\ref{fig2}. As expected, the most significant radial term is the spherical one, $V_0(R)$. By including radial terms with up to $\lambda_{\text{max}}=12$ in the Legendre series expansion of Eq. (\ref{eq_PES}), we ensured a relative error of the Legendre expansion of only 0.1~$\%$ with respect to the \emph{ab initio} PES.

For cold and ultracold collisions, the long-range tail of the PES plays a critical role for evaluating the scattering cross sections. Therefore, we chose a certain $R_{\textrm{mid}}$ for each $V_{\lambda}$ beyond which the radial coefficient was replaced by the long-range dipole--induced dipole interaction,


\begin{equation}
\label{eq_vl3}
V_{\lambda}(R\geq R_{\textrm{mid}})=\frac{C_6^{\lambda}}{R^6}
\end{equation}

\noindent
The values of $C_6^{\lambda}$ were obtained by fitting the numerical values from Eq.~(\ref{eq_vlambda}) between $R_{\textrm{mid}}$ and 30~\AA. Additionally, a polynomial interpolation was used to ensure the continuity of the potential and its first derivative at $R_{\textrm{mid}}$.

\section{Collision dynamics \label{Sec.Theory}}

In evaluating the quantum dynamics of the AlF+He collisions, we make use of the framework developed by Arthurs, Dalgarno, and Bates~\cite{Arthurs1960} for rigid rotor--atom collisions, omitting hyperfine structure. 

\subsection{State-to-state integral cross sections}
\label{sec:state_to_state}

The state-to-state integral scattering cross section for collisions leading from an initial state $j$ of the rotor to its final state $j'$ is given by
\begin{equation}
\label{eq-xs}
\sigma_{j\rightarrow j'} (E_k)= \frac{4\pi}{k^2(2j+1)}\sum_{Jll'}(2J+1)|T^{JM}_{j,l;j'l'}|^2        
\end{equation}
where $E_k$ is the collision energy, $k$ the wavenumber of the relative motion atom-molecule motion, $k^2=2\mu E_k$, with $\mu$ the reduced mass of the atom-molecule system, and $T^{JM}_{j,l;j'l'}$ is the transition matrix that encapsulates the transition probabilities between different scattering channels characterized by the rotational quantum numbers $j$ and $j'$ and partial waves $l$ and $l'$ before and after the collision. In the absence of external fields, only channels $(j,l)$ and $(j',l')$ are coupled that conserve the total angular momentum $J$ and its projection $M$ on the quantization axis (scattering coordinate $\bf{R}$).
Note that throughout this section, we use atomic units. 


In the center-of-mass frame, the Hamiltonian of the atom-molecule system is given by
\begin{equation}
\label{eq2}
    H=-\frac{1}{2\mu}\nabla^2_R+\frac{\hat{l}^2}{2\mu R^2}+V(R,\theta)+H_{\text{rot}}
\end{equation}
where the first term represents the radial kinetic energy operator along the scattering coordinate $\bf{R}$, the second term denotes the centrifugal term of the kinetic energy operator, $V(R,\theta)$ the atom-molecule PES, cf. Sec.~\ref{Sec.PES}, and $ H_{\text{rot}}$ the rotational Hamiltonian of the molecule. The last is given by  
\begin{equation}
    H_{\text{rot}}Y_j^{m_j}(\hat{r}) = [B_0j(j+1)-D_0(j(j+1))^2]Y_j^{m_j}(\hat{r})
\end{equation}
with $Y_j^{m_j}(\hat{r})$ the spherical harmonics, $B_0$ the rotational constant of the vibrational ground state of the molecule, $D_0$ is the centrifugal distortion constant, $j$ the rotational quantum number and $m_j$ its projection on the direction of the molecular axis, $\hat{r}=\frac{\bf{r}}{|\bf{r}|}$. 

The total energy of the rotor is comprised of its kinetic/collision energy and the internal energy of the fragment\footnote{A fragment represents the distinct energy levels of the system under consideration in the asymptotic region. Therefore, a fragment corresponds to many channels in the interaction region.} it corresponds to 
\begin{equation}
E_j=E_k+B_0j(j+1)-D_0(j(j+1))^2    
\end{equation}
which is an eigenvalue of the Schr\"odinger equation
\begin{equation}
\label{eq_scho}
    H\Psi^{JM}_{jl}=E_j\Psi^{JM}_{jl}
\end{equation}
In the coupled basis, preferable in the absence of external fields, the eigenfunctions of Eq.(\ref{eq_scho}) are given by
\begin{equation}
\label{eq_psi}
    \Psi^{JM}_{jl}=\sum_{j'l'}\frac{g_{jl,j'l'}^{JM}(R)}{R} I_{j'l'}^{JM}(\hat{r},\hat{R})
\end{equation}
with
\begin{equation}
 I_{jl}^{JM}(\hat{r},\hat{R})=\sum_{m_jm_l}C^{JM}_{jm_jlm_l} |jm_jlm_l\rangle    
\end{equation}
the angular part of the wave function that entails the coupling of the molecule-fixed coordinate $\hat{r}$ to the scattering coordinate $\hat{R}=\frac{\bf{R}}{|\bf{R}|}$ via the Clebsch-Gordan coefficients $C^{JM}_{jm_jlm_l}$.

By substituting Eq.(\ref{eq_psi}) into Eq.(\ref{eq_scho}) we obtain
\begin{eqnarray}
\label{eq-cc}
\bigg[ -\frac{d^2}{dR^2}+ k_{j'}^2  + \frac{l'(l'+1)}{R^2} \bigg] g_{jl,j'l'}^{JM} (R)  \nonumber  = \\  2\mu \sum_{j''l''}V^{J}_{j''l'';j'l'}(R)g_{jl,j''l''}^{JM}(R)
\end{eqnarray}
where 
\begin{eqnarray}
k_{j'}^2=2\mu \bigg[ E_k +B_0j(j+1) -D_0(j(j+1))^2 - \nonumber \\B_0j'(j'+1) +D_0(j'(j'+1))^2\bigg]    
\end{eqnarray}
is the wavenumber squared pertaining to the $(j,j')$ channel, and 
\begin{equation}
 V^{J}_{j'l';jl}(R) = \int \int I^{JM*}_{j'l'}(\hat{r},\hat{R})V(R,\theta) I^{JM}_{jl}(\hat{r},\hat{R})d\Omega_r   
\end{equation}
are the potential matrix elements.
By substituting from Eq.~(\ref{eq_PES}), we obtain
\begin{eqnarray}
\label{advantage}
V^{J}_{j'l';jl}(R) = \sum_{\lambda}^{\lambda_{\textrm{max}}} V_{\lambda}(R) (-1)^{J+l'+l} \nonumber \\
\sqrt{(2j+1)(2l+1)(2j'+1)(2l'+1)} \nonumber \\
\tj{l}{\lambda}{l'}{0}{0}{0} \tj{j}{\lambda}{j'}{0}{0}{0} \Gj{j}{\lambda}{j'}{l'}{J}{l}
\end{eqnarray}
with $(...)$ and $\{...\}$ the 3j and 6j symbols, respectively. 
Thus the simple form of Eq. (\ref{advantage}) results by virtue of the Legendre expansion of the PES, introduced in Sec. \ref{Sec.PES}.

The set of coupled differential equations (\ref{eq-cc}) is solved numerically. In the asymptotic region where the interaction potential in negligible compared to the collision energy, the numerical results can be matched to the expected analytic asymptotic behavior 
\begin{eqnarray}
g_{jl,j'l'}^{JM}(R \rightarrow \infty)\sim \delta_{jj'}\delta_{ll'}\sin{\left( k_{j'}R-\frac{l'\pi}{2}\right)} \nonumber \\
+\frac{e^{\imath (k_{j'}R-l'\pi/2)}}{\sqrt{k_{j'}}}  T^{JM}_{jl,j'l'}   
\end{eqnarray}
leading to the numerical transition matrix and thus, via Eq.(\ref{eq-xs}), to the state-to-state cross section.



\subsection{Diffusion and viscosity transport cross sections}
\label{sec:diffandvisc}

The diffusion and viscosity cross sections are defined by~\cite{Mott}
\begin{equation}
\label{eqdiff}
    \sigma_D(E_k)=\int \frac{d\sigma_{\textrm{el}}(E_k)}{d\Omega}(1-\cos{\theta})d\Omega
\end{equation}
and
\begin{equation}
\label{eqvisco}
    \sigma_\eta(E_k)=\int \frac{d\sigma_\textrm{el}(E_k)}{d\Omega}(1-\cos^2{\theta})d\Omega
\end{equation}
where $\frac{d\sigma_\textrm{el}(E_k)}{d\Omega}$ is the elastic differential cross section, $d\Omega=2\pi \sin \theta d \theta$ the solid angle element, and $\theta$ the scattering angle. Quantum mechanically, these cross-sections evaluate to~\cite{Mott}
\begin{equation}
\label{sig_D}
    \sigma_D(E_k)=\frac{4\pi}{k^2}\sum_{l=0}^\infty(l+1) \sin^2{[\delta_{l+1}(E_k)-\delta_l(E_k)]}
\end{equation}
and 
\begin{equation}
\label{sig_eta}
    \sigma_\eta(E_k)=\frac{4\pi}{k^2}\sum_{l=0}^\infty\frac{(l+1)(l+2)}{2l+3} \sin^2{[\delta_{l+2}(E_k)-\delta_l(E_k)]}
\end{equation}
respectively, wherein $\delta_l(E_k)$ is the phase-shift for a given partial wave $l$ and collision energy $E_k$.
The quantum elastic scattering cross-section expressed in terms of the phase shifts boils down to
\begin{equation}
\label{sig_el}
    \sigma_{\textrm{el}}(E_k)=\frac{4\pi}{k^2}\sum_{l=0}^\infty(2l+1) \sin^2\delta_{l}
\end{equation}

Classically, the transport cross-sections are given by~\cite{mason,Hirschfelder}
\begin{align}
\label{classical_sigma}
& \sigma^C_D(E_k)=2\pi\int_{0}^{\infty}\left[1-\cos{(\chi(E_k,b))}\right]b \,db \\
\label{classical_sigma2}
& \sigma^C_\eta(E_k)=2\pi\int_{0}^{\infty}\left[1-\cos^2{(\chi(E_k,b))}\right]b \,db
\end{align}
with $\chi$ the deflection angle and $b$ the impact parameter:



\begin{equation}
\label{deflection_angle}
\chi(E_k,b)=\pi-2b\int_{R_c}^{\infty} \frac{R^{-2}dR}{\sqrt{1-V_0(R)/E_k-b^2/R^2}}
\end{equation}
where $R_c$ is the distance of closest approach in the collision, obtained by solving
\begin{equation}
\label{close_approach}
1-V_0(R_c)/E_k-b^2/R_c^2=0
\end{equation}
Integration of $\chi$ over all impact parameters then yields $\sigma_D$ and $\sigma_\eta$ as given above by Eqs. (\ref{classical_sigma}) and (\ref{classical_sigma2}). 

The expressions for the transport cross sections can be  thermally averaged, resulting in 
\begin{equation}
   \sigma^C_D(T)=\frac{1}{2(k_BT)^3}\int_0^{\infty}\sigma_D(E_k) \exp\bigg[-\frac{E_k}{k_BT}\bigg] E_k^2dE_k
\end{equation}
and
\begin{equation}
\sigma^C_{\eta}(T)= \frac{1}{6(k_BT)^4}\int\sigma_\eta(E_k) \exp\bigg[-\frac{E_k}{k_BT}\bigg] E_k^3dE_k  
\end{equation}
with $k_B$ the Boltzmann constant and $T$ the temperature. These thermally averaged cross sections occur in the kinetic theory of gas transport \cite{Hirschfelder}.
 
\section{Computational details}
\subsection{Classical cross sections}
The classical cross-sections were calculated by solving  Eqs.~(\ref{classical_sigma}) to (\ref{close_approach}) numerically using Python. The integration was performed using scipy's integrate.quad command for an initial guess of the point of closest approach. Only the spherical component of the PES, $V_0(R)$, was taken into account, cf. Eqs. (\ref{deflection_angle}) and (\ref{close_approach}).

\begin{table}[]
\begin{tabular}{ll}
\hline
\hline
$\mu$ (a.m.u.) & 3.68207364173 \\ \hline
$B_0^{\textrm{AlF}}$ (cm$^{-1}$) &  0.5499923150168107 \\ 
$D_0^{\textrm{AlF}}$ (cm$^{-1}$) & 0.000001040719977 \\ 
$R_{\textrm{min}} (\textrm{\AA})$ & 1.6 \\
$R_{\textrm{max}} (\textrm{\AA})$ & 120 \\ 
$\textrm{DR} (\textrm{\AA})$ & 0.008  \\ 
OTOL $(\textrm{\AA}^2)$ & 0.001 \\ 
DTOL $(\textrm{\AA}^2)$ & 0.1 \\ 
NLEVEL & 12 \\ \hline \hline
\end{tabular}
\caption{Summary of parameters used for the quantum coupled-channel calculations. Here DR denotes the step-size for integration, OTOL and DTOL the off-diagonal and diagonal cross section tolerance thresholds for convergence, and NLEVEL the number of angular momentum quantum levels used to construct the basis set.}
\label{table1}
\end{table}

\subsection{Quantum elastic cross-sections}
The quantum elastic cross sections were obtained by solving the single-channel Schr\"odinger equation using Numerov's method. We employed 10$^5$ steps between 3.6~$a_0$ and $R_{\text{max}}$ (300~$a_0$ for the lowest collision energy and 60~$a_0$ for the highest) including an appropriate number of partial waves to guarantee a convergence better than 1$\%$. And, as in the case of the classical calculations, only the spherical component $V_0(R)$ of the potential was used.

\subsection{Quantum multi-channel calculations}
Throughout this study, we made use of the spectroscopic constants summarized in Table III of Ref.~\citep{Truppe2019} (i.e., $B_0 = 16488.3548$ MHz and $D_0 = 0.0312$ MHz) to calculate the rotational levels of the AlF rotor in its electronic and vibrational ground state. We used the MOLSCAT software package\citep{MOLSCAT, HUTSON2019} to perform quantum multi-channel scattering calculations of the rotationally inelastic cross-sections based on the coupled-channel approach of Dalgarno \textit{et al.} described in Sec.~\ref{Sec.Theory}. The coupled-channel equations were solved using the log-derivative method of Manolopoulos~\citep{LDMD_Manolopoulos} between 1.6$~\textrm{\AA}$ and 120$~\textrm{\AA}$ with a step size of 0.008$~\textrm{\AA}$. This method is an improved version of the Johnson's log-derivative algorithm \citep{JohnsonLD}, which propagates the log-derivative matrix $Y(R)=\Psi'(R)[\Psi(R)]^{-1}$ instead of the wave function. The maximum total angular momentum of the system required to reach convergence is automatically decided by the code by setting the parameter JTOTU $\geq 99999$, and is found not to exceed a maximum value of 50. In addition, all the multi-channel calculations comprised 12 rotational states with $j=0-11$. A summary of the parameters used for the scattering calculations is given in Table~\ref{table1}.

\section{Results and discussion}

\begin{figure}
\centering
\includegraphics[width=1\linewidth]{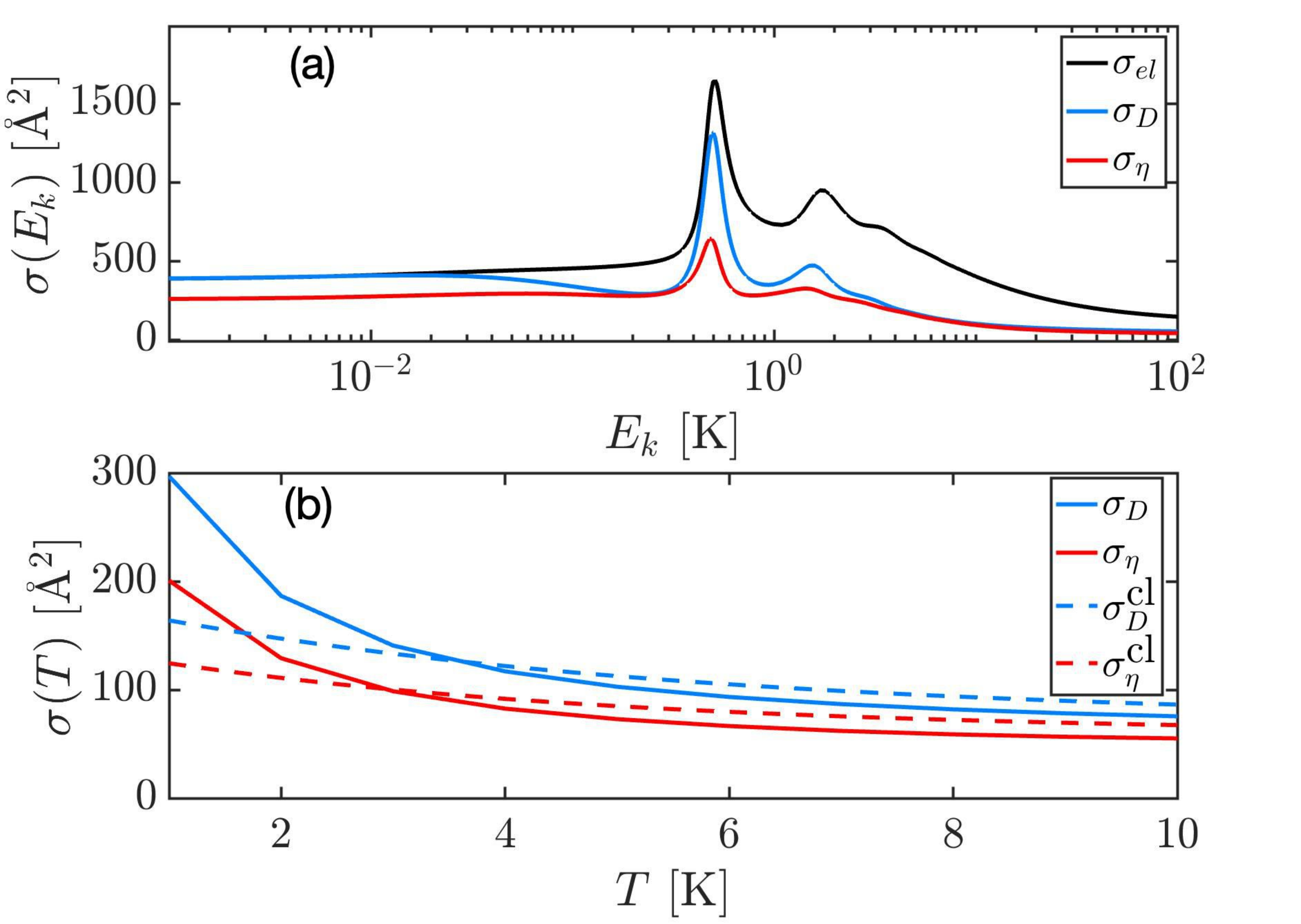}
\caption{Transport cross-sections for AlF in cryogenic helium. Panel (a) displays the elastic, diffusion and viscosity cross sections as functions of the collision energy. Cf. the $0\to 0$ elastic cross section in Fig.~\ref{fig5} (thick blue). Panel (b) displays the temperature-averaged quantum transport cross sections (full curves) on temperature. The corresponding classical cross sections are shown by dashed curves.}
\label{fig3}
\end{figure}

The calculated quantum elastic and transport cross-sections are shown in panel (a) of Fig.~\ref{fig3} as functions of collision energy. The cross-sections are seen to exhibit three shape resonances, which occur at collision energies between 0.5 and 10~K, which happens to be the energy range relevant for buffer-gas cooling experiments. At collision energies below~0.1 K, one can see the emergence of the Wigner threshold behavior. This is consistent with the value of the van der Waals energy~\cite{Jones2006,JPRBook}
\begin{equation}
    E_{\text{vdW}}=\left(\frac{2\hbar^6}{\mu^3C_6}\right)^{1/2}
\end{equation}
which, for the AlF+He system, evaluates to $E_{\text{vdW}}$=0.168~K, where we made use of the \emph{ab initio} value of the $C_6^0$ pertaining to the spherical component $V_0$ of the potential, $C_6^0=-24.6$~a.u.

In the ultracold regime which, for the system considered, sets in below 10 mK, the diffusion cross section is seen to converge to the elastic one, whereas the viscosity cross section converges to a limit of its own. This behavior is due to the s-wave scattering that dominates the ultracold collision regime. For s-wave scattering, the phase shift is related to the scattering length $a$ via
\begin{equation}
    \lim_{k \to 0} \delta_0(k)\rightarrow -ka
\end{equation}
By substituting this result into Eqs. (\ref{sig_D}), (\ref{sig_eta}), and  (\ref{sig_el}), we obtain
\begin{align}
 & \lim_{k \to 0}\sigma_D(E_k)\rightarrow 4\pi a^2 \nonumber \\
  & \lim_{k \to 0}\sigma_{\eta}(E_k)\rightarrow \frac{2}{3}4\pi a^2 \nonumber\\
  & \lim_{k \to 0}\sigma_{\textrm{el}}(E_k)\rightarrow 4\pi a^2
\end{align}
which confirms the numerically found convergence behavior of the diffusion and viscosity cross-sections displayed in panel (a) of Fig.~\ref{fig3}.

\begin{figure}
\includegraphics[width=1 \linewidth]{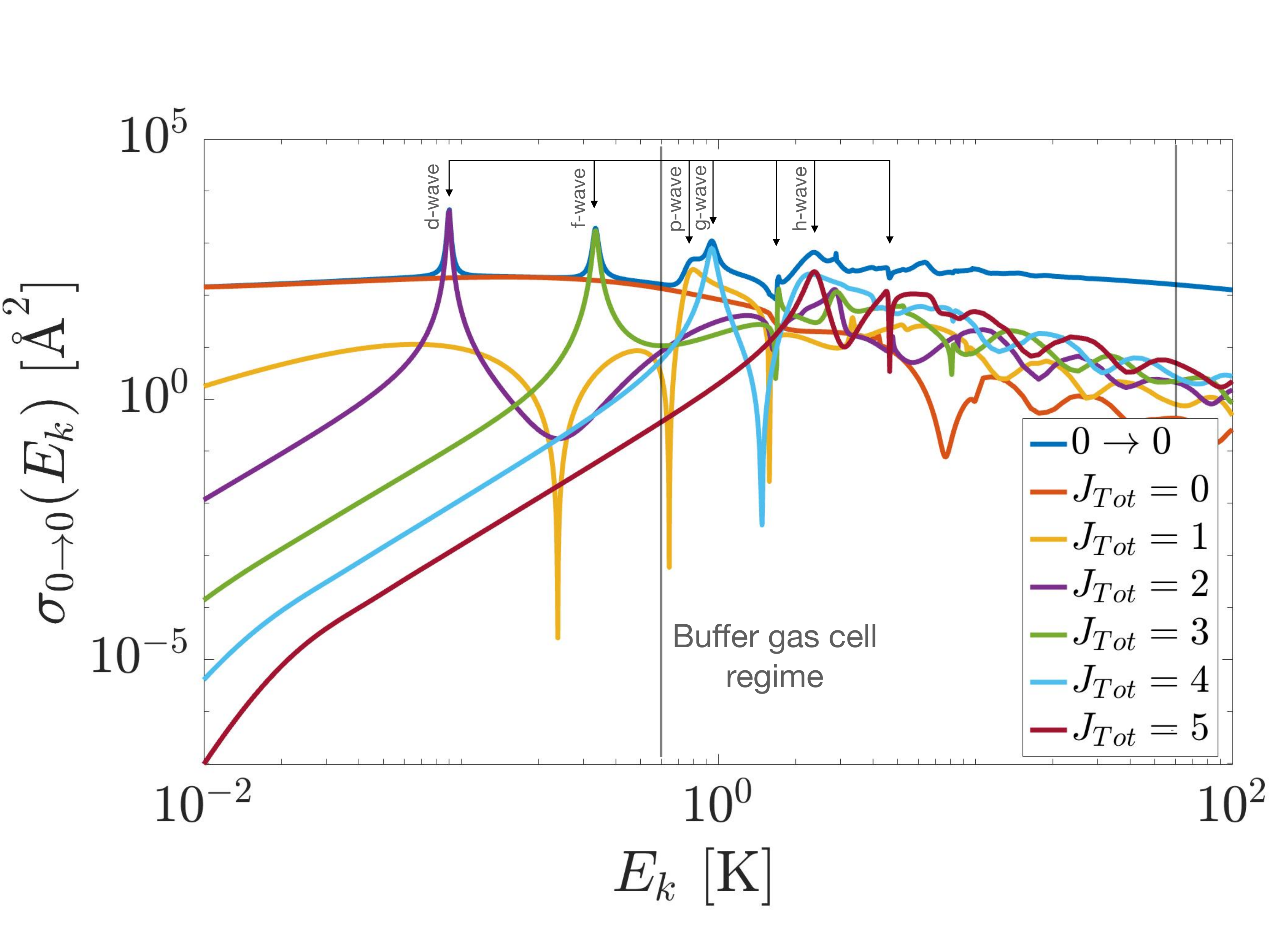}
\caption{Elastic cross-sections and partial wave contributions. The s-, p-, d-, f-, g- and h-wave partial contributions explain specific features of the elastic cross section curve for the ground state ($j=0$) of the AlF rigid rotor. The prominent shape resonances around $2.5$~K occur in a buffer gas cell experiment.} 
\label{fig4}
\end{figure}

A comparison between the classical and quantum temperature-averaged transport cross sections is shown in panel (b) of Fig.~\ref{fig3}, which reveals that both treatments predict similar trends in the temperature dependence. For higher temperatures, the classical and quantum results are in a better agreement than at lower temperatures, due to the washing out of contributions from a large number of partial waves. In the temperature range relevant for buffer-gas cooling experiments, i.e., $T\lesssim 10$~K, the discrepancy between the classical and quantum mechanical results becomes more significant -- and thus a quantum treatment of the transport cross-sections desirable. 

We have also calculated the classical thermally averaged diffusion cross section at 20~K in order to compare it with the previously reported results for YbF($^2\Sigma^+$) in He~\cite{Skoff2011} at that temperature. It turns out that the diffusion cross section of AlF in He is 70.13~\AA$^2$, which comes quite close to the value of 79.6~\AA$^2$ for YbF in He. This suggests that the transport cross sections for the two fluoride diatomics are essentially independent of either molecular symmetry ($^2\Sigma$ versus $^1\Sigma$) or atom properties (Yb versus Al).



A molecule injected into a buffer-gas cell undergoes also relaxation of its internal degrees of freedom, in particular of molecular rotation. This process, termed rotational quenching, is driven by rotationally inelastic collisions from a given initial rotational state. In what follows, we present the results of our calculations of the rotationally inelastic AlF+ He collisions that are based on the multi-channel coupled-channel approach described in Sec. \ref{sec:state_to_state}.


\begin{figure}
\includegraphics[width=1 \linewidth]{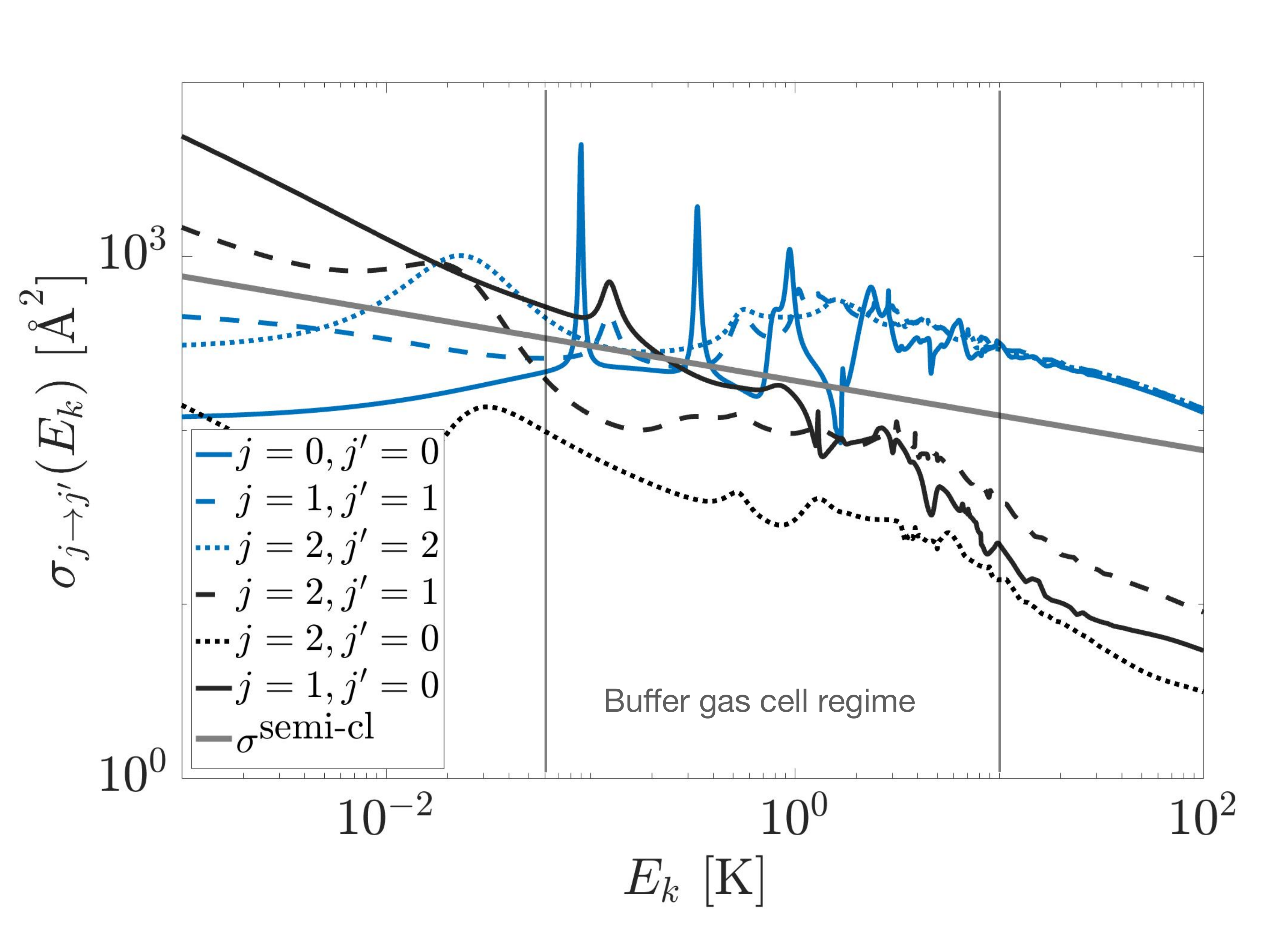}
\caption{Elastic and rotationally inelastic scattering cross-sections for the first three initial rotational states ($j=0,1,2$) of the AlF rotor. For collision energies greater than about $1.7$ K, elastic scattering begins to dominate over all inelastic scattering channels.}
\label{fig5}
\end{figure}

Inelastic channels affect the elastic one via the unitarity of the scattering matrix \cite{JPRBook}, p. 26, as can be gleaned from Fig.~\ref{fig4}, where we find a larger number of resonances than in the elastic cross-section, cf. panel (a) of Fig.~\ref{fig3}. In addition, the resonances are often less distinct than in the single-channel, elastic case because of their mutual proximity. These additional resonances come about for two main reasons: (1) availability of more channels and (2) the anisotropy of the PES. Having more channels increases the probability of resonances, provided the underlying atom-molecule interaction couples the channels. After performing a decomposition of the cross sections into partial waves, we were able to conclude that, barring two resonances at $0.09$~K and $0.33$~K, the s-wave contribution dominates at collision energies of up to 0.67 K. The remaining partial-wave contributions account for the first few prominent features of the elastic cross section. In particular, for collision energies $\gtrsim 10$~K, the different partial waves exhibit a rather oscillatory behavior that is averaged out once the partial waves are summed up to yield the elastic cross section.


\begin{figure}
\includegraphics[width=1 \linewidth]{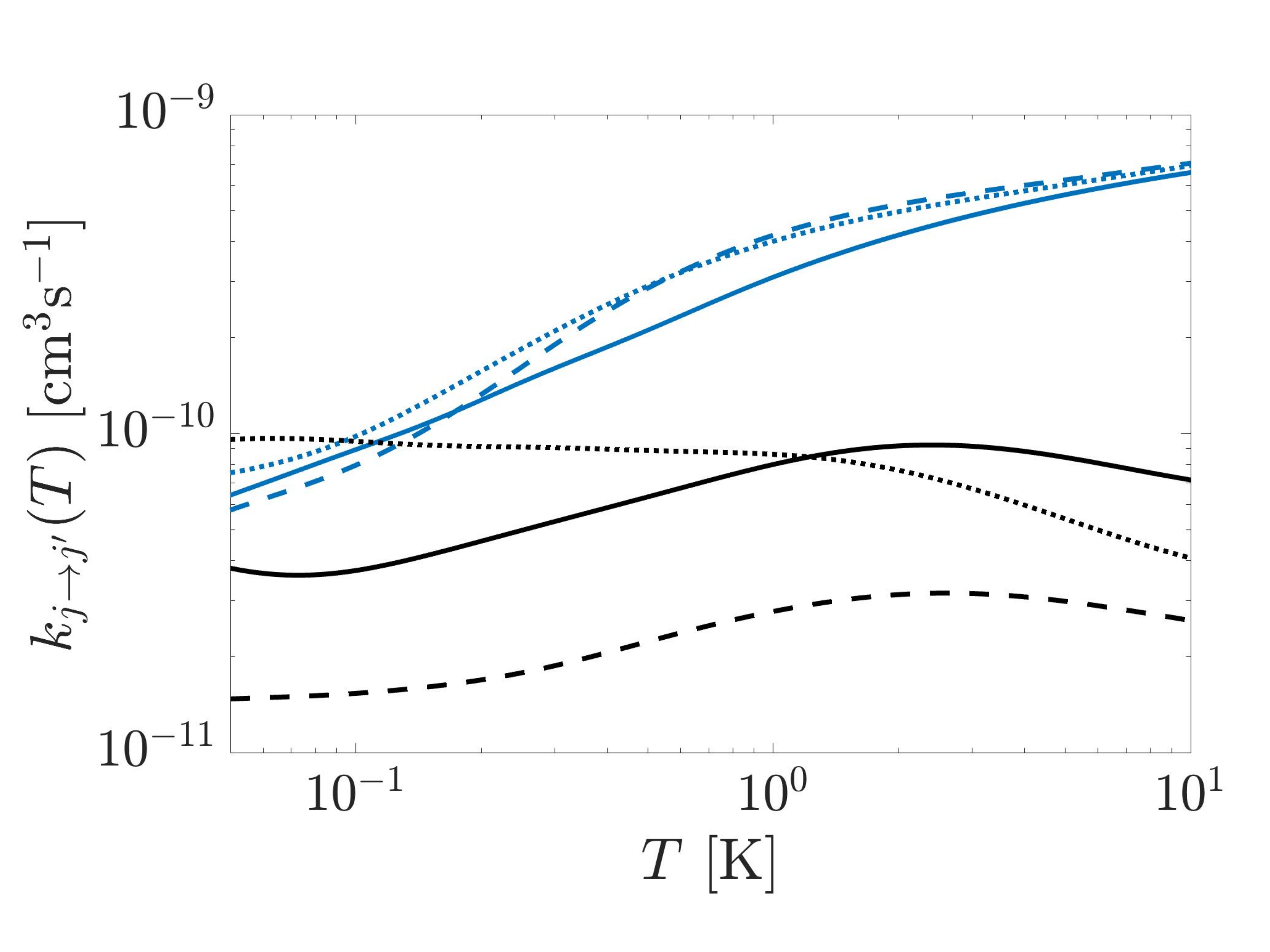}
\caption{Rate constants for elastic (blue curves) and rotationally inelastic (black curves) processes as functions of temperature. The elastic rates dominate over the inelastic ones beyond collision energies in excess of $0.1$ K. For detailed legend, see Fig.~\ref{fig5}.}
\label{fig6}
\end{figure}

In Fig.~\ref{fig5} we show both elastic and inelastic cross sections for the first three rotational states of AlF ($j=0,1,2$) as functions of collision energy. At low collision energies (the left-most regime), quantum behavior dominates the dynamics, leading to a pronounced state-dependent effect on the elastic and inelastic cross sections. For instance, the elastic cross section tends to a different scattering length depending on the rotational state considered. The positions of the resonances in both the elastic and inelastic cross sections depend significantly on the initial and final rotational states. The rotational-state dependence of the inelastic cross-sections follows in part from energetic considerations (the larger the energy gap between different initial and final states, the lower the transition probability). However, additional features of the system in question need to be included in order to rationalize its behavior: For instance, the inelastic cross-section $\sigma_{2\rightarrow 0}(E_k)$ is less than $\sigma_{1\rightarrow 0}(E_k)$ because the two processes are driven, respectively, by the $V_2(R)$ and $V_1(R)$ components of the potential and $V_2(R)<V_1(R)$, as revealed by our \emph{ab initio} PES calculations.

\begin{table}[]
\begin{tabular}{ccc}
\hline
\hline
System &  $T=5$~K & $T=10$~K \\ \hline
HF-He &  $5\times 10^{-12}$ & $5\times 10^{-12}$ \\ 
CO-He & $3.4\times 10^{-11}$ & $3.2\times 10^{-11}$ \\
AlF-He & $5.4\times 10^{-11}$ & $4\times 10^{-11}$ \\ \hline \hline 
\end{tabular}
\caption{State-to-state rate constants $k_{1\to 0}(T)$ in cm$^{3}$s$^{-1}$ for different $^1\Sigma^+$-He scattering systems. The values of the constants for the HF-He and CO-He systems have been taken from Refs.~\citep{HFHe, COHe}.}
\label{table2}
\end{table}


The transport rate constants of a fluid \cite{Hirschfelder} can be obtained from the transport cross sections introduced in Section~\ref{sec:diffandvisc}. In the absence of molecular vibration, the rotational state-to-state rate constants are given by \cite{Montero2014}, 
\begin{eqnarray}
k_{j\rightarrow j'}(T)=<\sigma_{j\rightarrow j'}(E_k)v>=\left(\frac{8}{\pi \mu \beta^3}\right)^{1/2} \nonumber \\
\int_{0}^{\infty}\sigma_{j\rightarrow j'}(E_k)\exp{(-E_k/\beta)}E_kdE_k    
\end{eqnarray}
with $\beta=k_BT$. The dependence of the state-to-state rate constants on temperature is displayed in Fig.~\ref{fig6} for a temperature range relevant to buffer-gas cooling. What we see is a large difference between the rates of elastic and inelastic scattering in the cryogenic buffer gas. Similarly as in the case of the cross sections, the rate constants for the exchange of the largest rotational quanta are suppressed. However, the elastic rate constants are almost independent of the rotational state, which reflects the average nature of the rate coefficient. 

\begin{figure}
\centering
\includegraphics[width=1 \linewidth]{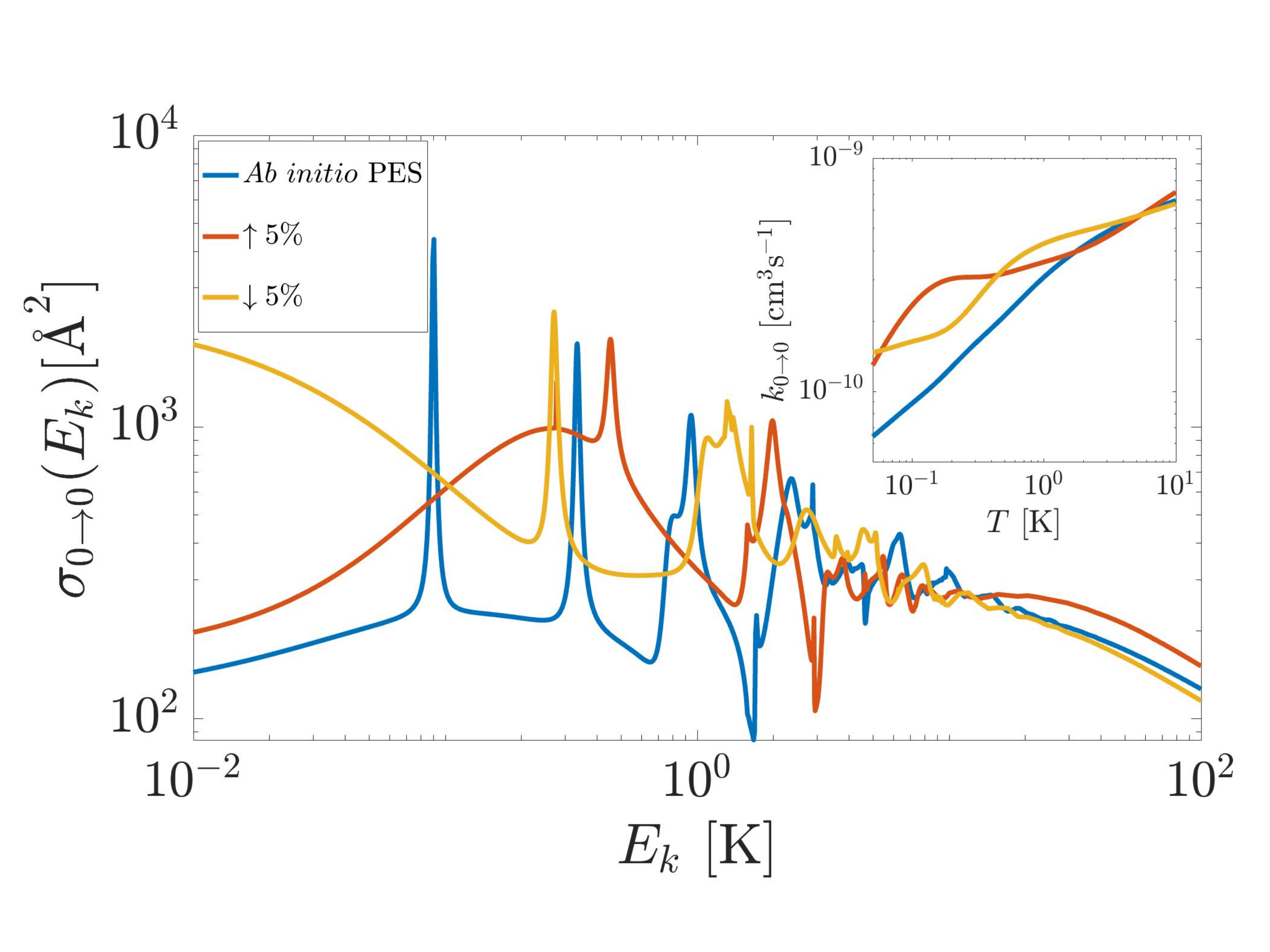}
\caption{Sensitivity to shifts in the PES up ($\uparrow$) or down ($\downarrow$) by 5\%. The density of resonances essentially remains unchanged within the collision energy regime accessible to buffer gas experiments. Inset shows rate coefficients for rotational quenching as a function of temperature.}
\label{fig7}
\end{figure}

Fig.~\ref{fig6} also reveals that the state-to-state rate constants corresponding to a relaxation of the rotational degrees of freedom of the molecule are only weakly temperature-dependent. A similar behavior has been reported for other $^1\Sigma^+$-He collision systems, as summarized in Table~\ref{table2}. The more efficient relaxation of AlF in He can be rationalized by invoking the adiabaticity parameter, $\xi=\tau_{\textrm{coll}}/\tau_{\textrm{rot}}\propto B_0/\sqrt{\mu}$, where $\tau_{\textrm{coll}}$ and $\tau_{\textrm{rot}}$ are, respectively, the collision time and rotational period of the molecule, see also Table \ref{table1}. For $\xi \gg 1$, the collision time exceeds the rotational period and hence the collision proceeds adiabatically, without changing the rotational state of the molecule. On the other hand, for $\xi\ll 1$ the collision process is nonadiabatic, leading to efficient rotational energy transfer. Therefore, heavier molecules with smaller rotational constants, such as AlF compared with CO or HF,   lead to small values of $\xi$ and thus a more nonadiabatic behavior (more facile rotational quenching). Based on the same argument, we conclude that the MgF-He system will show relaxation rates similar to those of the AlF-He system, but that BaF-He will exhibit a more efficient quenching due to BaF's smaller rotational constant~\cite{database}.


Given that the inaccuracy of the {\it ab initio} quantum chemistry calculations exceeds
the collision energies of $\lesssim$100~K that are at play in cryogenic helium cooling, we tested how sensitive the elastic scattering cross sections are to changes of the PES. To this end, the PES was either raised or lowered throughout by $5\%$ and the calculations repeated. The error thus simulated overestimates the inaccuracies of our method in the long-range tail of the potential but may underestimate them in the short-range region. The simulations are displayed in Fig.~\ref{fig7}, which attests that the positions and the widths of the resonances are affected and how. Nevertheless, the density of the resonances remains unchanged. The simulation also reveals that in
the thermal regime, for energies greater than 10 K, the cross sections are less sensitive to changes in the PES, as expected. The inset of Fig.~\ref{fig7} shows the effects of varying the PES on the state-to-state rate constants, which are seen to remain qualitatively unaffected.




\section{Conclusions}
We undertook a study of the collisions between aluminum monofluoride (AlF) molecules and helium (He) atoms at collision energies relevant to buffer-gas cooling by cryogenic helium. Our study, based on an accurate {\it ab initio} potential energy surface and quantum multichannel scattering theory, revealed high thermalization rates for both translational and rotational degrees of freedom of the molecule, on the order of  $10^{10}$ and $10^{11}$ cm$^3$s$^{-1}$, respectively. 

The large anisotropy of the AlF-He PES (due to the preferred He-F attraction over that between He and Al) and a small rotational constant of AlF translates into a complex resonance structure in elastic and rotationally inelastic scattering cross sections. However, at collision energies $\gtrsim 10$~K, the quantum effects are found to average out and most of the scattering observables become independent of the rotational state of the molecule. Classical calculations of the transport cross sections are found to be valid up to temperatures of $5$~K with a relative error $\lesssim$ $15\%$. 

Although possible inaccuracies of  the PES may affect the positions of the resonances, their density was found to remain the same. Moreover, we showed that the state-to-state rate constants are quite robust against inaccuracies of the PES for temperatures relevant to buffer-gas cell experiments. These findings confirm AlF as a serious candidate for efficient sympathetic cooling.


Our results on the AlF-He system are of a more general interest. For one, our classical transport cross sections benchmark the accuracy of such methods for $^1\Sigma^+$ molecule-He collisions. Furthermore, considering the  adiabaticity parameter, we estimate that MgF will be quenched as efficiently as AlF. Finally, owing to its smaller rotational constant, we expect BaF to be quenched more efficiently than AlF. 



\section{Acknowledgments}
MK acknowledges support by the IMPRS for Elementary Processes in Physical Chemistry. BF thanks John Doyle and Hossein Sadeghpour for their hospitality during his stay at the Harvard Department of Physics and at the Harvard-Smithsonian Institute for Theoretical Atomic, Molecular, and Optical Physics (ITAMP). 

\bibliography{Scattering}
\end{document}